\documentclass[aps,prl,twocolumn,superscriptaddress,showpacs,showkeys]{revtex4}

\usepackage{dcolumn}
\usepackage{amsmath}
\usepackage{graphicx}
\usepackage{rotating}
\usepackage{multirow}

\begin{document}

\newlength{\figurewidth}
\ifdim\columnwidth<10.5cm
  \setlength{\figurewidth}{0.95\columnwidth}
\else
  \setlength{\figurewidth}{10cm}
\fi
\setlength{\parskip}{0pt}
\setlength{\tabcolsep}{6pt}
\setlength{\arraycolsep}{2pt}

\title{Scaling and universality in proportional elections}

\author{Santo Fortunato}
\affiliation{School of Informatics, Indiana University, Bloomington, IN 47406, USA}
\affiliation{Complex Networks Lagrange Laboratory (CNLL), ISI Foundation, Torino, Italy}

\author{Claudio Castellano}
\affiliation{INFM unit\`a di Roma 1 and SMC, Dipartimento di
Fisica, Universit\`a "La Sapienza", P.le A. Moro 2, 00185 Roma, Italy}

\begin{abstract}

A most debated topic of the last years is whether 
simple statistical physics models can 
explain collective features of social dynamics. 
A necessary step in this line of endeavour is 
to find regularities in data referring to large scale 
social phenomena, such as scaling and universality.
We show that, in proportional elections,
the distribution of the number of votes received by candidates
is a universal scaling function, identical in different countries and years.
This finding reveals the existence in the voting process of a 
general microscopic dynamics that does not depend on the historical,
political and/or economical context where voters operate.
A simple dynamical model for the behaviour of voters, similar to a
branching process, reproduces the universal distribution.

\end{abstract}



\maketitle

Many social nontrivial phenomena emerge spontaneously
out of the mutual influence of a large number of
individuals~\cite{ball,applause, Panic,jams}, similarly
to large-scale thermodynamic behavior resulting from the interaction
of a huge number of atoms or molecules. 
However, human interactions are neither purely mechanical
nor reproducible, both typical requirements for a physical 
description of a process.
Nevertheless the collective behavior of large groups of individuals
may be independent of the details of social interactions and
individual psychological attributes, and be instead the consequence of generic
properties of the elementary interactions, allowing for a simple
'statistical physics' modeling.

In this spirit, microscopic models have been recently proposed to
account for collective social phenomena, like
the formation of consensus on a specific
topic~\cite{Deffuant, Sznajd, Krapivsky},
the creation of common cultural traits and their
dissemination~\cite{Axelrod}, the origin and evolution of
language~\cite{Nowak,Stauffer_rev}, etc.
While models are studied quantitatively in great detail,
the comparison with real-world social phenomena is often
merely qualitative and on this account it is not possible to make
a real discrimination between competing models.
This in turn limits their predictive power, making it unclear
if there is at all a gain in the understanding of social dynamics through
statistical physics. 

Elections are an ideal playground for a quantitative validation of the
approach to social dynamics inspired by physics.
They constitute a precise global
measurement of the state of the opinions of the electorate.
A large number of individuals are involved and
big datasets are available for many countries, thus allowing accurate
quantitative investigations. 

In this paper we present compelling evidence that 
elections data display properties of more traditional physical 
phenomena characterized by collective behavior and self-organization, 
i.e. scaling and universality. We show that, in proportional 
elections, the distribution of the number of votes received by candidates   
is universal, i.e. it is the same function in different countries and years,
when the number of votes is rescaled according to the strength of
the party each candidate belongs to.
We claim that the universal voting behavior is due to  
the spreading of the word of mouth from
the candidate to the voters, which we model as
a sort of branching process involving the acquaintances of a candidate. 

Early studies revealed that the histogram of the fraction $\nu$ of voters
supporting a candidate  within a constituency in Brazilian parliamentary
elections is described by a $1/\nu$ law,
in the central part of the range of
the variable $\nu$~\cite{CostaFilho,CostaFilho2,bernardes}.
A successive analysis of Indian elections~\cite{Gonzalez}
found a similar yet different histogram, hinting that the
distribution of the fraction of votes $\nu$ may exhibit some degree
of universality.
We have performed the same analysis on German, French, Italian and Polish
elections~\cite{Fortunato}, finding marked differences between
the various countries: the $1/\nu$ pattern is not general.

This lack of universality is a consequence of the fact that the number
of votes a candidate receives is the combination of two 
distinct factors: how many of the total number of electors vote for
the candidate's party and the personal appeal of the candidate within
the restricted pool of voters for his/her party.
The first factor strongly depends on policy-related issues: typically voters
know the position of all parties with respect to the political issues they
deem more relevant and they select the party that best matches their personal
views.
The second factor is instead practically independent from political issues.
Since candidates of the same party mostly share a common set of opinions
on ethical, social and economical issues,
the selection of a specific candidate has not to do with such issues, rather
it depends on a ``personal'' interaction
between the candidate and the voters. Typically voters know at most a few
of the candidates in their party list, and in this small subset they select
the one they will support.
Successful candidates are those able to establish some form of direct
or indirect contact with many potential voters during the electoral campaign.
This type of opinion dynamics is likely to give rise to universal phenomena.
The histogram of the total fraction $\nu$ of votes may conceal 
the actual regularities due to the voter dynamics,
as it entangles the two factors: a very popular candidate of a small party 
can have the same number of votes (but for completely different reasons)
of a relatively unpopular candidate of a very large party.

In the following we focus on the second factor, how the electors of
each party select candidates in their party list.
This rules out systems based on single member constituencies,
where every party/coalition presents a single candidate in each electoral
district, 
as well as proportional elections with closed lists, where voters are
not allowed to express preferences among party candidates: in this case the ranking
of candidates of a party is predetermined by the party.

The most suitable elections to investigate the elementary voter dynamics are
proportional elections with multiple-seat constituencies and open lists. 
In this electoral system, the country is divided in districts, and each
of them allocates a certain number of seats, $Q_{max}$,
typically between $10$ and $30$.
Within each district, each party $l$ presents a list of $Q_l \le Q_{max}$
candidates.
Voters choose one of the parties and also express their preference
among the candidates of the selected party.
Each party gets $n_l$ of the total number of seats,
in proportion to the number of votes it has received in the district.
The $n_l$ most voted candidates of party $l$ are elected.
In this way, the party plays no role
as to which of its candidates will be eventually elected, 
their success depending only on the free choice of voters.

We have considered three countries with such type of
electoral system: Italy (until 1992), Poland and Finland.
We use publicly available~\cite{Sitiweb} data sets for three elections
in Italy ($1958$, $1972$, $1987$), one in Poland ($2005$) and one in
Finland ($2003$).
The total number of  candidates ranges from $2,029$ for the Finnish 
elections in $2003$ to $10,658$ for the Polish elections in $2005$.

To factor out the policy-related role of the parties, we
keep track, for candidate $i$ that receives $v_i$ votes, also of two
other parameters: $Q_{l_i}$, i.e. the
number of candidates of the party list $l_i$, where $i$ belongs, and 
$N_{l_i}$, total number of votes collected by the $Q_{l_i}$ candidates of list $l_i$.

The distribution of the number of votes collected by candidates is in general
a function of the three variables $P(v,Q,N)$.
We show instead that $P(v,Q,N)$ is actually a function of a single rescaled variable.
\begin{figure}
\begin{center}
\includegraphics[width=8cm]{Figure2A.eps}
\includegraphics[width=8cm]{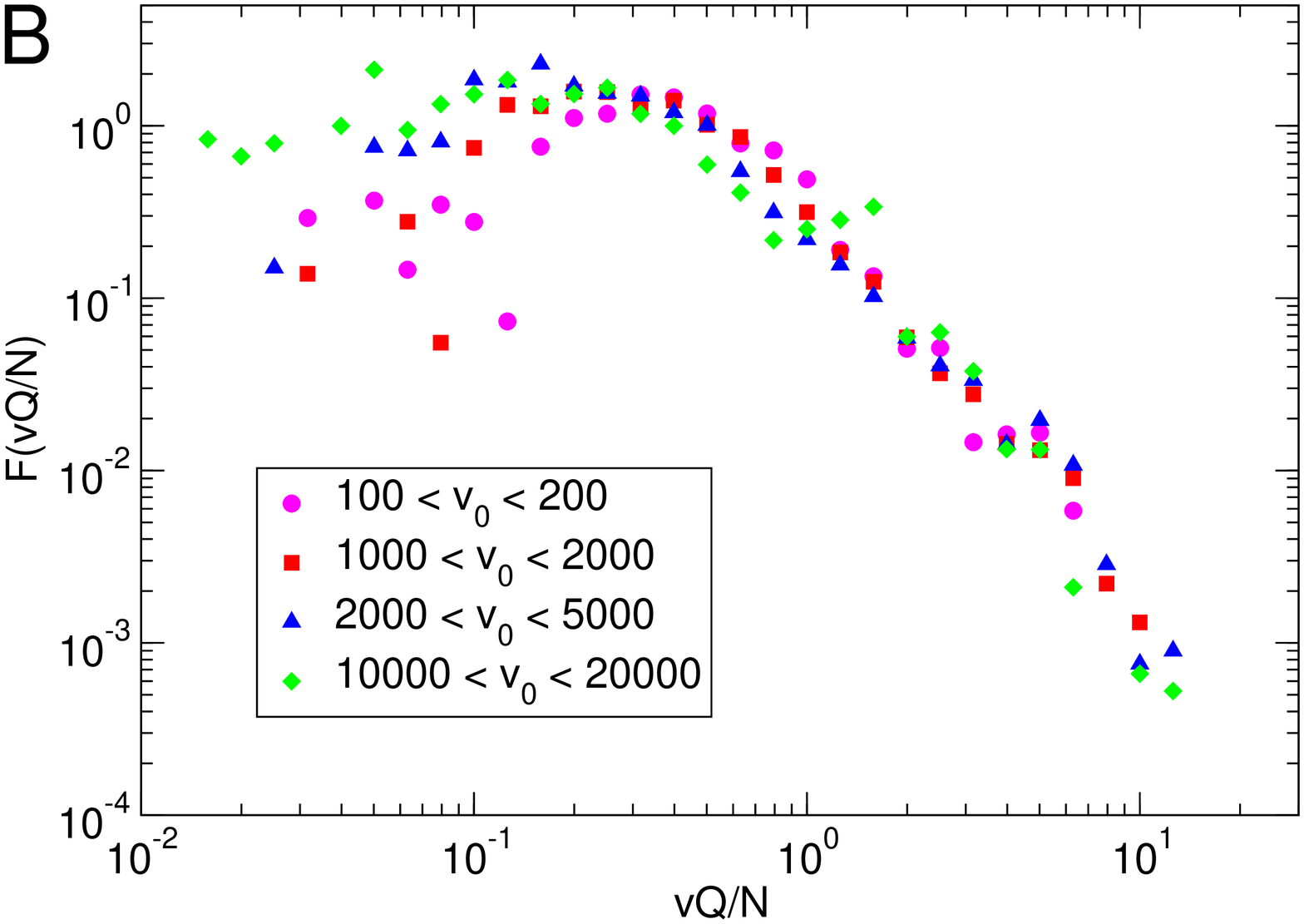}
\caption {\label{Figure2}A.
Scaling behavior of the distribution of votes received by candidates.
Data refer to the Italian parliamentary elections in 1972,
but we obtained very similar results from the analysis of each dataset.
The histogram $P(v,Q,N)$ only depends on the ratio $v_0=N/Q$, so $P(v,Q,N)=P_0(v,v_0)$.
B. The function $P_0(v,v_0)$ shown in A only depends 
on the ratio $v/v_0=vQ/N$.
Data refer to the Italian parliamentary elections in 1972.}
\end{center}
\end{figure}
We start by showing that $P(v,Q,N)$ does not depend on $N$ and $Q$ separately,
but only on the ratio $v_0=N/Q$,
which is the average number of votes collected
by a candidate in his/her list. 
The curves of Fig.~\ref{Figure2}A  correspond to three different values of $v_0$.
Since $v_0$ is a continuous variable,
fixing $v_0$ actually means selecting those lists with values of $v_0$
within a narrow range. 
For each value of $v_0$ we fix a threshold for the total number $N$
of votes and  further filter the data by separating the lists with
$N$ larger/smaller than the threshold.
For a fixed $v_0$, the resulting histograms are the same for both data
samples, proving that the distribution $P(v,Q,N)$ is actually only a
function of the arguments $v$ and $v_0$, $P_{0}(v,v_0)$.

But a close inspection of the function $P_0(v,v_0)$ reveals
that the dependence on two variables is actually only apparent:
the distribution of the rescaled variable $v/v_0=vQ/N$ turns out
to be independent of $v_0$.
Again, we filter the data by putting together candidates belonging to
lists such that the ratio $v_0=N/Q$ falls in one of four narrow windows.
For each set of candidates we derive the histogram of the rescaled
variable $v/v_0$: the four curves have a remarkable overlap,
so there is no dependence on $v_0$ but only on $v/v_0$ (Fig.~\ref{Figure2}B).
We conclude that $P(v,Q,N)$  
is actually a function of the single variable $vQ/N$, i.e. 
\begin{equation}
P(v,Q,N)=F(vQ/N).
\label{Eq1}
\end{equation}
Since $v_0=N/Q$ is the average number of votes collected by a 
candidate in his/her list, 
the ratio $vQ/N=v/v_0$ is an index of the performance of a candidate 
against his/her competitors in the same list. If $v/v_0 < 1$, the candidate
has received less votes than average; if $v/v_0 \gg 1$, he/she performed
much better than average.

\begin{figure}
\includegraphics[width=8cm]{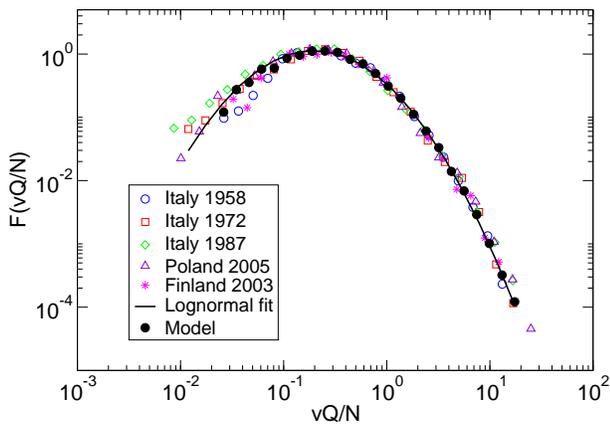}
\caption {\label{Figure3}
Universality of the scaling function $F(vQ/N)$ across different
countries and years. The lognormal fit, performed on the Polish curve,
describes very well the data. The universal curve is well reproduced by 
our model, where the dynamics of the voters' opinions reflects the spreading
of the word of mouth in the party's electorate.}
\end{figure}

Eq.~\ref{Eq1} indicates that each election can be characterized by a
single function $F(vQ/N)$.
A comparison between the scaling functions $F$ for all five data sets is
presented in Fig.~\ref{Figure3} and gives an even more striking result:
the scaling function $F(vQ/N)$ is the same for different countries and years.
The universal curve is very well reproduced by a {\it lognormal} function, i.e.
\begin{equation}
F(vQ/N)=\frac{N}{\sqrt{2\pi}\sigma vQ}e^{-{(\log(vQ/N)-\mu)^2}/2\sigma^2},
\label{Eq2}
\end{equation}
with $\mu=-0.54$, $\sigma^2=-2\mu=1.08$.
The relation $\sigma^2=-2\mu$ is due to the fact that the expected value
of the variable $vQ/N=v/v_0=1$ and that the expected value of a lognormal
distributed variable is $\exp(\mu+\sigma^2/2)$.

The universality of the distribution $F(vQ/N)$ is truly remarkable. 
The elections considered span a period of thirty years, in which
deep cultural, economic and social transformations have occurred: there is no
hint of that  in the data pattern. Likewise, differences between countries as
diverse as Italy, Poland and Finland do not play any role.
This calls for a modelization in terms of simple mechanisms of interaction
between voters (and candidates), regardless of the details of the social,
cultural and economic environment. 

\begin{figure}
\includegraphics[width=8cm]{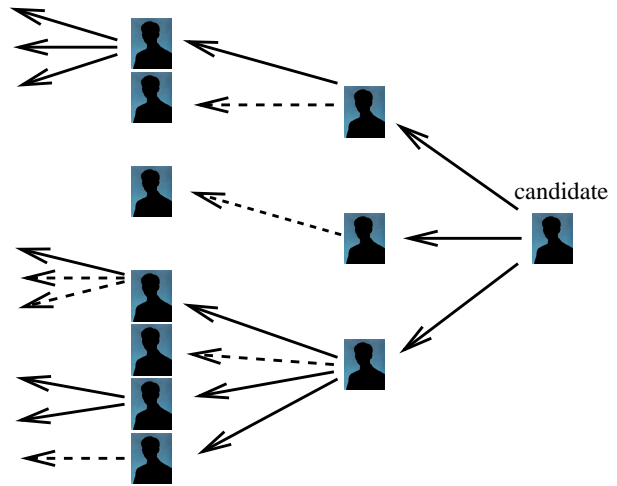}
\caption {\label{Figure4}
Spreading of the word of mouth among voters.
The candidate (right) convinces some of his/her contacts to vote
for him/her. The convinced voters become ``activists'' and try
to convince some of their acquaintances, and so on.
Successful interactions are indicated by solid lines,
unsuccessful interactions are displayed as dashed lines.}
\end{figure}

The spreading of word of mouth is known to be a very effective   
vehicle of diffusion of new products among potential
buyers~\cite{word-of-mouth0}.
We interpret the electoral results using a simple opinion dynamics model
based on word of mouth: electors that have already chosen a candidate try to
convince their peers to vote for the same candidate (Fig.~\ref{Figure4}).
At the beginning, only candidates have an opinion (they vote for themselves).
The dynamics starts with the candidates trying to convince their acquaintances.
The people convinced by each candidate become
activists and in turn try to convince their contacts to vote for their
candidate, and so on. Only undecided voters can be convinced.
Not all interactions result in an undecided voter being convinced:
persuasion occurs only with probability $r$. 
Models of opinion spreading with similar features have been introduced
recently~\cite{Moreira, Travieso}.

We implement the process by representing the electorate of a party
as a set of tree-like communities of voters, with
candidates as roots, as shown schematically in Fig.~\ref{Figure4}.
We have as many independent trees as candidates, and each
candidate acts on the nodes of its own tree, representing the voters
within its sphrere of influence, and not on the others. 
The distribution $p(k)$ of the number $k$ of contacts of a voter 
has to be broad, as there are very active people
that try to convince as many voters as possible, as well as less active
ones, that do not feel particularly
involved or motivated.
We assume therefore that $p(k)$ is described by a power law, i.e. that the 
probability $p(k)$ that a voter has $k$ acquaintances is
$p(k)\sim k^{-\alpha}$, with $\alpha>1$.
To completely fix the distribution
$p(k)$, we fix the lower bound of $k$, that we indicate with $k_{min}$.

Every iteration of the process consists
in the persuaded voters trying to convince their undecided contacts,
each with probability $r$. One keeps track of the running number of
convinced voters, which increases with time. 
The process stops when this number equals $N$, where $N$ is the size of
the electorate of that party in the constituency.  
Our model is similar to a collection of branching processes evolving in
parallel, coupled via the condition that they stop when the total number
of convinced voters reaches $N$.
Branching processes have a large number of applications in the physics
literature, from modeling of forest fires~\cite{forest}, 
to percolation~\cite{mezhlumian},
to self-organized criticality~\cite{zapperi}. 
It is important to stress, however, that our model is not
a usual branching process, but a full-fledged new process, with different
and nontrivial properties. The key point is that, while in
branching processes once a node has decided how to branch it remains
frozen, here convinced voters keep trying to persuade their contacts:
branching events can occur at any point in the trees.

We study the dynamics of our model by means of computer simulations.
For each choice of the 
parameters $\alpha$, $k_{min}$ and $r$, that we consider, we repeat the 
process several times. Each time we store the number of votes received by
every candidate. When enough scores are collected, the histogram $P(v,Q,N)$
is determined.

The distribution $P(v,Q,N)$,
obtained via numerical simulations, exhibits the scaling properties
of the empirical distribution, i.e. it obeys Eq.~\ref{Eq1}.
In Fig.~\ref{Figure3} we fit the model distribution
to the empirical curve. To account for
finite size effects, we ran the simulations on the same set of
values for $Q$ and $N$ that occur in the empirical datasets. and convoluted 
the resulting curves. The model curve of Fig.~\ref{Figure3} is the
convolution of  the distributions obtained from each pair of $Q$ and $N$,
for $\alpha=2.45$, $k_{min}=10$ and $r=0.25$:
the agreement is remarkable.  

The histogram $F(vQ/N)$ depends rather slowly on the three model
parameters $\alpha$, $k_{min}$ and $r$; besides, 
the decreasing part of the curve is very robust~\cite{Fortunato}.

We have shown that election data reveal impressive regularities when 
the role of policy-related issues is factored out so that the voter dynamics 
only relies on the contact of the candidates with the voters.
This pattern of behavior is the same in different countries and times 
and hence is affected neither by individual
features of the voters nor by the environment where the voters live.
We conclude that the underlying voting dynamics is elementary and can
be described by simple statistical physics models.  
A branching-like process representing the propagation
of word of mouth reproduces the universal distribution of votes
for candidates. 
We expect this universality to hold for other countries where the electoral
system is (or will be in the future) proportional with open lists.

As a potential application of our results, since the relative performance
of a candidate in a list has the same distribution everywhere,
the index $vQ/N$ is an objective estimate of the popularity of a candidate,
independently of the
constituency and the year of the election; this gives parties an unbiased
quantitative basis to decide internal rankings and hierarchies.

Word of mouth spreading is a crucial ingredient to explain other instances of 
collective social dynamics, such as the spreading of news and fads in a
population and the diffusion of new products among potential consumers.
From the analysis of these processes other
signatures of universality may emerge. This research direction
may strengthen the confidence on the applicability of statistical physics
to explain large scale social dynamics.

\begin{acknowledgments}
We thank C. Brooks, A. Flammini, M. Marsili, 
D. Stauffer and A. Vespignani for useful discussions.
\end{acknowledgments}

\end{document}